\documentclass[aps, prd, onecolumn,showpacs, showkeys, preprintnumbers, preprint, nofootinbib, superscriptaddress]{revtex4-2}
\usepackage[sort&compress]{natbib}
\usepackage{multirow}
\usepackage{graphicx}
\usepackage{amssymb,amsbsy,amsmath,amsfonts}
\usepackage[latin1]{inputenc}

\begin{document}
\title{Exotic states with triple charm}

\author{M. Bayar}
\email{melahat.bayar@kocaeli.edu.tr}
\affiliation{Department of Physics, Kocaeli Univeristy, 41380, Izmit, Turkey.}
\affiliation{Departamento de F\'isica Te\'orica and IFIC, Centro Mixto Universidad de Valencia-CSIC, Institutos de Investigaci\'on de Paterna, Aptdo. 22085, 46071 Valencia, Spain.}

\author{A. Mart\'inez Torres}
\email{amartine@if.usp.br}
\affiliation{Universidade de Sao Paulo, Instituto de Fisica, C.P. 05389-970, Sao 
Paulo, Brazil.}
\affiliation{Departamento de F\'isica Te\'orica and IFIC, Centro Mixto Universidad de Valencia-CSIC, Institutos de Investigaci\'on de Paterna, Aptdo. 22085, 46071 Valencia, Spain.}

\author{K. P. Khemchandani}
\email{kanchan.khemchandani@unifesp.br}
\affiliation{Universidade Federal de Sao Paulo, C.P. 01302-907, Sao Paulo, Brazil.}
\affiliation{Departamento de F\'isica Te\'orica and IFIC, Centro Mixto Universidad de Valencia-CSIC, Institutos de Investigaci\'on de Paterna, Aptdo. 22085, 46071 Valencia, Spain.}

\author{R. Molina}
\email{Raquel.Molina@ific.uv.es}
\affiliation{Departamento de F\'isica Te\'orica and IFIC, Centro Mixto Universidad de Valencia-CSIC, Institutos de Investigaci\'on de Paterna, Aptdo. 22085, 46071 Valencia, Spain.}

\author{E. Oset}
\email{Eulogio.Oset@ific.uv.es}
\affiliation{Departamento de F\'isica Te\'orica and IFIC, Centro Mixto Universidad de Valencia-CSIC, Institutos de Investigaci\'on de Paterna, Aptdo. 22085, 46071 Valencia, Spain.}

\begin{abstract}
In this work we investigate the possibility of the formation of states from the dynamics involved in the $D^*D^*D^*$ system by considering that two $D^*$'s generate a $J^P=1^+$ bound state, with isospin 0, which has been predicted in an earlier theoretical work. We solve the Faddeev equations for this system within the fixed center approximation and find the existence of $J^P=0^-$, $1^-$ and $2^-$ states with charm $3$, isospin $1/2$, masses $\sim 6000$ MeV, which are manifestly exotic hadrons, i.e., with a multiquark inner structure.
\end{abstract}



\maketitle
\date{\today}

\section{Introduction}
The discovery of $T_{cc}$ by the LHCb collaboration~\cite{LHCb:2021vvq,LHCb:2021auc} in the $DD\pi$ invariant mass distribution is a turning point in the field of hadron spectroscopy, showing the existence of a state with doubly open charm flavor content, thus, clearly exotic in the sense that it does not qualify as a conventional $q\bar q$ meson. While other exotic states, the $X_0(2866)$ and $X_1(2900)$, containing $c$ and $s$ open flavors, have been found before~\cite{LHCb:2020bls,LHCb:2020pxc}, this is the first time that the discovery of a doubly charm meson is being reported experimentally.~The nature of $T_{cc}$ as a $D^*D$ bound state, decaying to $DD\pi$, has found a generalized support~\cite{Li:2021zbw,Meng:2021jnw,Feijoo:2021ppq,Wu:2021kbu,Ling:2021bir,Yan:2021wdl,Huang:2021urd,Xin:2021wcr,Fleming:2021wmk,Ren:2021dsi,Chen:2021cfl,He:2021smz,Dong:2021bvy,Ling:2021bir,Wu:2021kbu}. Correspondingly, the $D^*D^*$ system has also been studied from this point of view in Refs.~\cite{Li:2012ss,Dong:2021bvy,Albaladejo:2021vln,Du:2021zzh,Dai:2021vgf}. It should be pointed out that predictions for both the $D^*\bar K^*$ and $D^*D^*$ exotic molecular states had already been made earlier in Ref.~\cite{Molina:2010tx}.

The existence of exotic states with charm 2 raises the question on whether exotic states with higher open charm content, like charm 3, can be formed in Nature, for example, by adding a $D^*$ to the $D^*D^*$ system. The topic of three body systems made with mesons has captured attention recently. A review of different states studied can be found in Table 1 of Ref.~\cite{MartinezTorres:2020hus}. A status report and prospects of multi-meson molecules is presented also in Ref.~\cite{Wu:2022ftm}. In this latter work, an observation is made worth stressing here: what differentiates ordinary nuclei from multi-meson aggregates is essentially the baryon conservation number that prevents the decay of nuclei into other nuclei with smaller baryon number. There is no meson number conservation and multi-meson states can decay to other states with fewer mesons, to the point that the large width would make the states unrecognizable as the meson number increases. Yet, it is surprising that in the study of multi-rho states done in Ref.~\cite{Roca:2010tf}, up to six $\rho$ mesons could be put together and the resulting states could be associated with the existing states $f_2(1270)$, $\rho_3(1690)$, $f_4(2050)$, $\rho_5(2350)$ and $f_6(2510)$, the latter one already with a very large width. However, although there is no meson number conservation, the flavor of quarks is conserved in strong interactions, which, in the context of the present work, means that a system with $ccc\bar q\bar q\bar q$ quarks ($q=u,d$) formed from three mesons cannot decay to a system with fewer mesons. Thus, if a state is found in this three meson system its width could be small. It is then conceivable that multi-meson states with multiple open flavor quantum numbers (omitting the $q\bar q$ pairs of the same flavor that can annihilate) could be relatively stable. We present here the case of the $D^*D^*D^*$ system that we find indeed bound, with a relatively small width.

Systems of three mesons with triple charm have been studied in Ref.~\cite{Luo:2021ggs} assuming a $D^*T_{cc}$ configuration. More concretely, the $D^*D^*D$ system is studied in Ref.~\cite{Wu:2021kbu} and the $D^*D^*D^*$ system in Ref.~\cite{Luo:2021ggs}, using the one boson exchange model for the interaction and solving the three-body Schr\"odinger equation with the Gaussian expansion method. We use instead the fixed center approximation (FCA) to the Faddeev equations that has been used to study many systems~\cite{MartinezTorres:2020hus}. We take advantage of the work of Ref.~\cite{Dai:2021vgf} where bound states of $D^*D^*$ are studied using an extension of the local hidden gauge approach of Refs.~\cite{Bando:1987br,Harada:2003jx,Meissner:1987ge,Nagahiro:2008cv} to the heavy quark sector, exchanging vector mesons, and the system is found more bound than the $T_{cc}$ as a $D^*D$ state. In the FCA one must choose a cluster of two particles and in this case we naturally take the bound $D^*D^*$ system, and a third particle, the other $D^*$, collides repeatedly with the components of the cluster. This is done in analogy to what was done in Ref.~\cite{Roca:2010tf} to study multi-rho states. The accuracy of the method to study three body systems of the type studied here has been shown in the recent work of Ref.~\cite{Wei:2022jgc} studying the $D\bar D K$ system, where similar results are obtained as in Ref.~\cite{Wu:2020job} using the Gaussian expansion method. We also obtain results in qualitative agreement with Ref.~\cite{Luo:2021ggs}, with some differences which are attributable to differences in the input used for the $D^*D^*$ interaction, as we discuss in Sec.~\ref{res}. In particular, we find bound states with isospin $I=1/2$, spin-parity $J^P=0^-$, $1^-$, $2^-$, out of which the $0^-$ state is more bound than the other two and has a larger strength in the three-body scattering matrix.

\section{Formalism}
\begin{figure}
\centering
\includegraphics[width=0.63\textwidth]{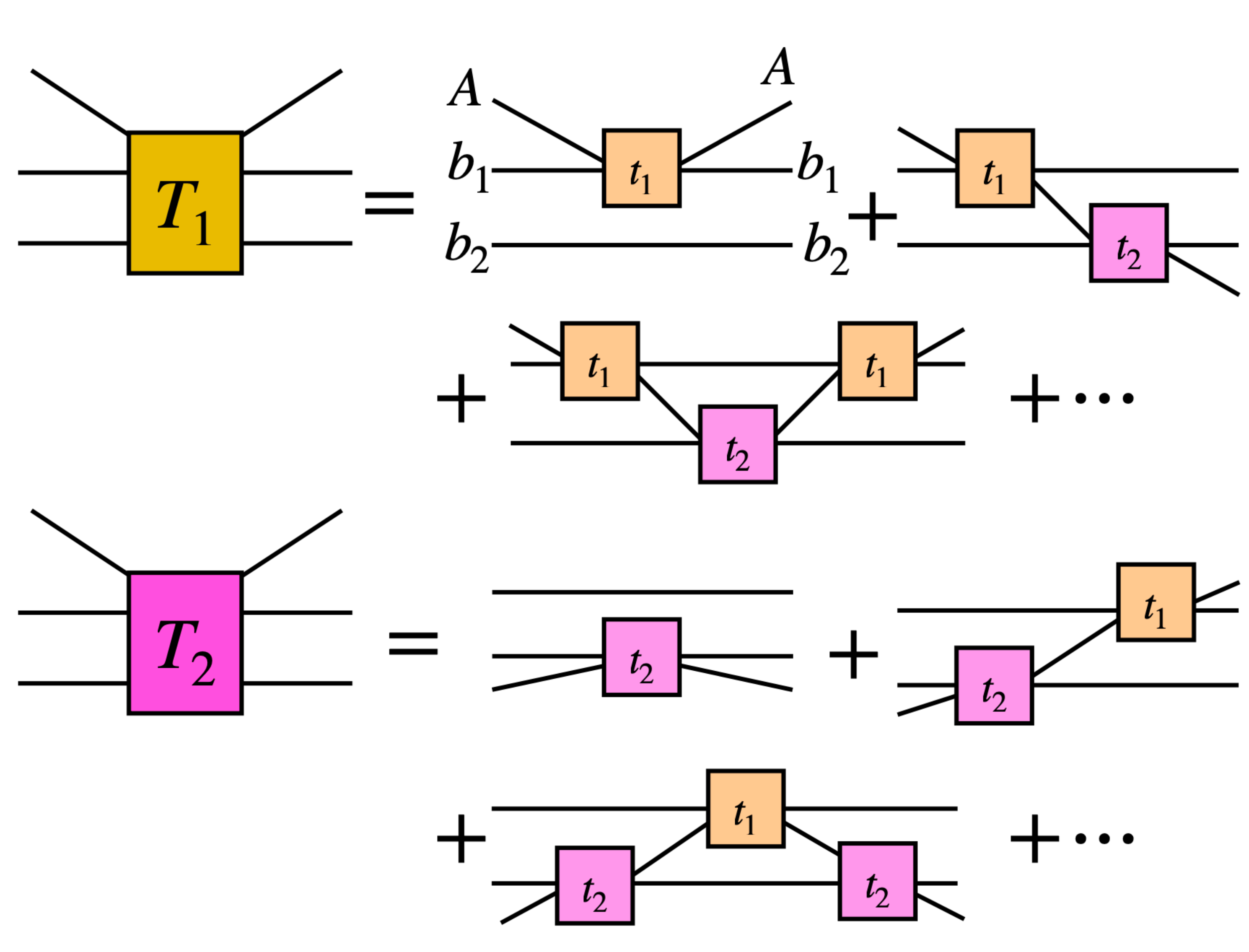}
\caption{Diagrams contributing to $T_1$ and $T_2$.}\label{FCA}
\end{figure}

In our approach, we determine the three-body $T$-matrix for the $D^*D^*D^*$ system and study the formation of states from its energy dependence on the real axis. To do this, we solve the Faddeev equations~\cite{Faddeev:1960su} within the FCA~\cite{Foldy:1945zz,Brueckner:1953zz,MartinezTorres:2020hus}. Such an approximation is feasible in this case since, as found in Refs.~\cite{Molina:2010tx,Dai:2021vgf}, the $D^*D^*$ interaction is attractive in nature and forms a bound state in $I=0$ with $J^P=1^+$, width of $\simeq$ 29 MeV and a binding energy\footnote{See the erratum for Ref.~\cite{Dai:2021vgf}.} of around 4-6 MeV. Thus, the interaction between the three particles of the system can be effectively considered as that of a $D^*$ with a cluster of isospin 0 and $J^P=1^+$ of the other two $D^*$'s, as shown in Fig.~\ref{FCA}. Since the $D^*$ interacting with the cluster can rescatter with any of the other two $D^*$'s of the cluster, we have the following set of coupled equations to determine the scattering matrix $T$ of the system~\cite{MartinezTorres:2020hus}:
\begin{align}
T_1&=t_1+t_1 G_0 T_2,\nonumber\\
T_2&=t_2+t_2 G_0 T_1,\label{T12}
\end{align}
where $T_i$, $i=1$, $2$, represents the contributions to the scattering matrix in which a particle $A$ (in this case, $D^*$)  rescatters first with the particle $b_i$ (a $D^*$ too) of the cluster $B$. In this way, the scattering matrix $T$ of the system is given by
\begin{align}
T=T_1+T_2.
\end{align}
In case of the system under investigation, i.e., $D^*D^*D^*$, it is clear that $T_1=T_2$. 

In Eq.~(\ref{T12}), $G_0$ represents the propagator of the particle $A$ in the cluster $B$, and it is given by~\cite{MartinezTorres:2020hus}
\begin{align}
G_0=\frac{1}{2M_B}\int\limits\frac{d^3 q}{(2\pi)^3}\frac{F(\vec{q})}{(q^{0})^2-\omega^2_A(\vec{q})+i\epsilon},\label{G0}
\end{align}
with $q^0$ being the  on-shell energy of particle $A$ in the $B$ rest frame, i.e.,
\begin{align}
q^0=\frac{s-m^2_A-M^2_B}{2M_B},\label{q0rest}
\end{align}
where $\sqrt{s}$ is the center-of-mass energy of the three-body system, $M_B$ is the mass of the cluster, and $\omega_A(\vec{q})=\sqrt{\vec{q}^{\,2}+m^2_A}$ is the energy related to the particle $A$ propagating in the cluster. In Eq.~(\ref{G0}), $F(\vec{q})$ is a form factor associated with the wave function of the particles forming the cluster~\cite{MartinezTorres:2020hus},
\begin{align}
F(\vec{q})&=\frac{1}{N}\int\limits_{|\vec{p}|,\,|\vec{p}-\vec{q}|<q_\text{max}}d^3p f(\vec{p})f(\vec{p}-\vec{q});\quad N=\int\limits_{|\vec{p}|<q_\text{max}}d^3p f^2(\vec{p});\nonumber\\
f(\vec{p})&=\frac{1}{\omega_{b_1}(\vec{p})\omega_{b_2}(\vec{p})}\frac{1}{M_B-\omega_{b_1}(\vec{p})-\omega_{b_2}(\vec{p})+i\epsilon},\label{FF}
\end{align}
with $\omega_{b_1(b_2)}$ being the energy of the particle $b_1$ ($b_2$) and $N$ is a normalization factor such that $F(\vec{q}=0)=1$. The value $q_\text{max}$ used in Eq.~(\ref{FF}) corresponds to the cut-off considered when regularizing the loops present in the Bethe-Salpeter equation in the study of the $D^*D^*$ system~\cite{Dai:2021vgf}. In Ref.~\cite{Dai:2021vgf}, three different cut-offs where considered, $q_\text{max}=450$, $550$ and $650$~MeV, and we will study the uncertainty that this range of $q_\text{max}$ produces in the three-body $T$-matrix. The factor $1/(2M_B)$ in Eq.~(\ref{G0}) is a normalization factor whose origin lies in the normalization of the fields when comparing the scattering matrix $S$ of a three-body system in which particle $A$ rescatters off particles $b_1$ and $b_2$ of the cluster with that where particle $A$ interacts with particle $B$~\cite{MartinezTorres:2020hus}. As a consequence of the normalization of these $S$-matrices, a normalization factor needs to be included in the kernels $t_i$, $i=1,2$, as well as in $G_0$. In particular, 
\begin{align}
t_i\to \frac{M_B}{M_{b_i}}t_i.\label{norm}
\end{align}

The kernels $t_i$, $i=1,2$, in Eq.~(\ref{T12}) [which include the normalization factor given in Eq.~(\ref{norm})] are combinations of two-body $Ab_i\to Ab_i$ $t$-matrices and describe the interaction of particle $A$ with particle $b_i$ for a given isospin and spin of the three-body system. To obtain $t_i$ we proceed as follows: in the isospin basis, we have the particles $b_1$ and $b_2$, of isospin $1/2$ each, forming a cluster with isospin $I_B=0$, i.e.,
\begin{align}
|I_B=0,I_{Bz}=0\rangle&=\frac{1}{\sqrt{2}}\Bigg[\Big|I_{b_1}=\frac{1}{2},I_{b_1z}=\frac{1}{2}\Big\rangle\otimes\Big|I_{b_2}=\frac{1}{2},I_{b_2z}=-\frac{1}{2}\Big\rangle\nonumber\\
&\quad-\Big|I_{b_1}=\frac{1}{2},I_{b_1z}=-\frac{1}{2}\Big\rangle\otimes\Big|I_{b_2}=\frac{1}{2},I_{b_2z}=\frac{1}{2}\Big\rangle\Bigg].
\end{align}
Next, we have the particle $A$, of isospin $I_A=1/2$, together with a cluster of isospin $I_B=0$, thus, the $AB$ system has isospin $I_{AB}=\frac{1}{2}$. In this way,
\begin{align}
\Big|I_{AB}=\frac{1}{2},I_{ABz}=\frac{1}{2}\Big\rangle=\Big|I_A=\frac{1}{2},I_{Az}=\frac{1}{2}\Big\rangle\otimes|I_B=0,I_{Bz}=0\rangle.\label{Isos}
\end{align}
It should be noted that calculating the right-hand side of Eq.~(\ref{Isos}) is not as straight forward as it may seem at a first glance. This is because the combination must be written in terms of the isospin of the $A-b_1$ system or in terms of the isospin of the $A-b_2$ system, depending on whether we calculate the kernel $t_1$ or $t_2$, respectively. In this way, to get, for example, $t_1$, we write the ket $|I_{AB},I_{ABz}\rangle$ as
\begin{align}
&\Big|I_{AB}=\frac{1}{2},I_{ABz}=\frac{1}{2}\Big\rangle_{Ab_1}=\frac{1}{\sqrt{2}}\Bigg[\Bigg(\Big|I_A=\frac{1}{2},I_{Az}=\frac{1}{2}\Big\rangle\otimes\Big|I_{b_1}=\frac{1}{2},I_{b_1z}=\frac{1}{2}\Big\rangle\Bigg)\nonumber\\
&\quad\otimes\Big|I_{b_2}=\frac{1}{2},I_{b_2z}=-\frac{1}{2}\Big\rangle-\Bigg(\Big|I_A=\frac{1}{2},I_{Az}=\frac{1}{2}\Big\rangle\otimes\Big|I_{b_1}=\frac{1}{2},I_{b_1z}=-\frac{1}{2}\Big\rangle\Bigg)\nonumber\\
&\quad\otimes\Big|I_{b_2}=\frac{1}{2},I_{b_2z}=\frac{1}{2}\Big\rangle\Bigg],
\end{align}
where the subscript on $|I_{AB},I_{ABz}\rangle$ indicates that we express the ket in terms of the isospin of the $A-b_1$ system. Since the results obtained for the three-body $T$-matrix of the system do not depend on the total isospin projection, we consider $I_{ABz}=1/2$. In this way, 
\begin{align}
&\Big|I_{AB}=\frac{1}{2},I_{ABz}=\frac{1}{2}\Big\rangle_{Ab_1}=\frac{1}{\sqrt{2}}\Bigg[|I_{Ab_1}=1,I_{Ab_1z}=1\rangle\otimes\Big|I_{b_2}=\frac{1}{2},I_{b_2z}=-\frac{1}{2}\Big\rangle\nonumber\\
&\quad-\frac{1}{\sqrt{2}}\Big(|I_{Ab_1}=1,I_{Ab_1z}=0\rangle+|I_{Ab_1}=0,I_{Ab_1z}=0\rangle\Big)\otimes\Big|I_{b_2}=\frac{1}{2},I_{b_2z}=\frac{1}{2}\Big\rangle\Bigg].\label{IAB1}
\end{align}

Once we have determined the isospin state related to the $AB$ system, we focus on the angular momentum part. In the angular momentum basis, we have a particle $A$ of spin $s_A=1$ interacting with a cluster $B$ of spin $s_B=1$.  The cluster $B$ is formed from the s-wave interaction of two particles, $b_1$ and $b_2$, of spins $s_{b_1}=s_{b_2}=1$, thus we have orbital angular momentum 0 for the cluster. We consider the interaction between particles $A$ and $b_i$ in s-wave, as done in Refs.~\cite{Molina:2010tx,Dai:2021vgf}. This means that the angular momentum of the $AB$ system, $j_{AB}$, as well as that of the $A-b_i$ systems, $j_{Ab_i}$, coincide with the corresponding spins, i.e., $s_{AB}$, $s_{Ab_i}$, respectively. 

Let us consider, for example, the case $j_{AB}=s_{AB}=1$ to illustrate the evaluation of the kernel $t_1$. Taking into account the spin related to each of the particles and using Clebsch-Gordan coefficients, we can write
\begin{align}
|s_{AB}=1,s_{ABz}=1\rangle&=\frac{1}{\sqrt{2}}\Big(|s_{A}=1,s_{Az}=1\rangle\otimes|s_{B}=1,s_{Bz}=0\rangle\nonumber\\
&\quad-|s_{A}=1,s_{Az}=0\rangle\otimes|s_B=1,s_{Bz}=1\rangle\Big),\label{sszAB}
\end{align}
where we have chosen the spin state with projection $s_{ABz}=1$ since the results do not depend on this choice. Once again, we need to determine the interaction of the $AB$ system in terms of that between $A$ and the constituents of the cluster $B$. Thus, it is required to decompose the state in Eq.~(\ref{sszAB}) in terms of the spin of particle $A$ combined with each of the constituents of $B$. Considering
\begin{align}
|s_B=1,s_{Bz}=0\rangle&=\frac{1}{\sqrt{2}}\Big[|s_{b_1}=1,s_{b_1z}=1\rangle\otimes|s_{b_2}=1,s_{b_2z}=-1\rangle\nonumber\\
&\quad-|s_{b_1}=1,s_{b_1z}=-1\rangle\otimes|s_{b_2}=1,s_{b_2z}=1\rangle\Big],\nonumber\\
|s_B=1,s_{Bz}=1\rangle&=\frac{1}{\sqrt{2}}\Big[|s_{b_1}=1,s_{b_1z}=1\rangle\otimes|s_{b_2}=1,s_{b_2z}=0\rangle\nonumber\\
&\quad-|s_{b_1}=1,s_{b_1z}=0\rangle\otimes|s_{b_2}=1,s_{b_2z}=1\rangle\Big],\label{sB}
\end{align}
we can write now the ket $|s_{AB}=1,s_{ABz}=1\rangle$ in terms of the spin of the $A-b_1$ or $A-b_2$ systems depending on whether we are interested in finding the kernel $t_1$ or $t_2$, respectively. In the former case, we have
\begin{align}
&|s_{AB}=1,s_{ABz}=1\rangle_{Ab_1}=\frac{1}{2}\Bigg[\Big(|s_{A}=1,s_{Az}=1\rangle\otimes|s_{b_1}=1,s_{b_1z}=1\rangle\Big)\otimes|s_{b_2}=1,s_{b_2z}=-1\rangle\nonumber\\
&\quad-\Big(|s_{A}=1,s_{Az}=1\rangle\otimes|s_{b_1}=1,s_{b_1z}=-1\rangle-|s_{A}=1,s_{Az}=0\rangle\otimes|s_{b_1}=1,s_{b_1z}=0\rangle\Big)\nonumber\\
&\quad\otimes|s_{b_2}=1,s_{b_2z}=1\rangle-\Big(|s_{A}=1,s_{Az}=0\rangle\otimes|s_{b_1}=1,s_{b_1z}=1\rangle\Big)\otimes|s_{b_2}=1,s_{b_2z}=0\rangle\Bigg],
\end{align}
finding then
\begin{align}
&|s_{AB}=1,s_{ABz}=1\rangle_{Ab_1}=\frac{1}{2}\Bigg[|s_{Ab_1}=2,s_{Ab_1z}=2\rangle\otimes|s_{b_2}=1,s_{b_2z}=-1\rangle\nonumber\\
&\quad+\Bigg(\frac{1}{\sqrt{6}}|s_{Ab_1}=2,s_{Ab_1z}=0\rangle-\frac{1}{\sqrt{2}}|s_{Ab_1}=1,s_{Ab_1z}=0\rangle-\frac{2}{\sqrt{3}}|s_{Ab_1}=0,s_{Ab_1z}=0\rangle\Bigg)\nonumber\\
&\quad\otimes|s_{b_2}=1,s_{b_2z}=1\rangle-\frac{1}{\sqrt{2}}\Big(|s_{Ab_1}=2,s_{Ab_1z}=1\rangle-|s_{Ab_1}=1,s_{Ab_1z}=1\rangle\Big)\nonumber\\
&\quad\otimes|s_{b_2}=1,s_{b_2z}=0\rangle\Bigg].\label{sAB1}
\end{align}
Once we have obtained the isospin and angular momentum parts of the state related to the $AB$ system, the ket characterizing it (written in terms of the isospin and angular momentum of the $Ab_i$ system) is given by
\begin{align}
\Big|I_{AB}=\frac{1}{2},I_{ABz}=\frac{1}{2};s_{AB},s_{ABz}\Big\rangle_{Ab_i}=\Big|I_{AB}=\frac{1}{2},I_{ABz}=\frac{1}{2}\Big\rangle_{Ab_i}\otimes|s_{AB},s_{ABz}\rangle_{Ab_i}.
\end{align}
The kernel $t_i$ can be obtained for a given angular momentum of the $AB$ system (which, as mentioned earlier, coincides with $s_{AB}$, with $s_{AB}=0,1,2$) and isospin of the $AB$ system, which in this case is $I_{AB}=1/2$, as
\begin{align}
t^{(I_{AB},s_{AB})}_i={}_{Ab_i}\Big\langle I_{AB}=\frac{1}{2},I_{ABz}=\frac{1}{2};s_{AB},s_{ABz}\Big|t_{Ab_i}\Big|I_{AB}=\frac{1}{2},I_{ABz}=\frac{1}{2};s_{AB},s_{ABz}\Big\rangle_{Ab_i}.\label{tAB}
\end{align}
For example, using Eqs.~(\ref{IAB1}) and (\ref{sAB1}), we have from Eq.~(\ref{tAB}),
\begin{align}
t^{(1/2,1)}_1=\frac{1}{16}\Big[5 t^{(1,2)}_{Ab_1}+3t^{(1,1)}_{Ab_1}+4t^{(1,0)}_{Ab_1}+\frac{5}{3}t^{(0,2)}_{Ab_1}+t^{(0,1)}_{Ab_1}+\frac{4}{3}t^{(0,0)}_{Ab_1}\Big],
\end{align}
where $t^{(I_{Ab_1},s_{Ab_1})}_{Ab_1}$ represents the two-body $t$-matrix describing the s-wave transition $Ab_1\to Ab_1$ with isospin $I_{Ab_1}$ and spin $s_{Ab_1}$. In particular, since $A$ and $b_1$ are $D^*$'s, we have
\begin{align}
t^{(1/2,1)}_1=\frac{1}{16}\Big[5 t^{(1,2)}_{D^*D^*}+3t^{(1,1)}_{D^*D^*}+4t^{(1,0)}_{D^*D^*}+\frac{5}{3}t^{(0,2)}_{D^*D^*}+t^{(0,1)}_{D^*D^*}+\frac{4}{3}t^{(0,0)}_{D^*D^*}\Big].\label{tIS1}
\end{align}

We can repeat this procedure for $s_{AB}=0,\,2$, finding
\begin{align}
t^{(1/2,0)}_1&=\frac{1}{4}\Big[3 t^{(1,1)}_{D^*D^*}+t^{(0,1)}_{D^*D^*}\Big],\nonumber\\
t^{(1/2,2)}_1&=\frac{1}{16}\Big[9 t^{(1,2)}_{D^*D^*}+3t^{(1,1)}_{D^*D^*}+3t^{(0,2)}_{D^*D^*}+t^{(0,1)}_{D^*D^*}\Big].\label{tIS02}
\end{align}
For the $D^*D^*D^*$ system, the particle $b_2$ is also a $D^*$, and the expression obtained for $t^{(I_{AB},s_{AB})}_2$ coincides with that of $t^{(I_{AB},s_{AB})}_1$. These latter $t$-matrices depend on the invariant mass of the $Ab_i$ cluster, which can be determined in the $B$ rest frame as
\begin{align}
s_{Ab_i}=\Big(p_A+\frac{1}{2}p_B\Big)^2&=m^2_A+\frac{1}{4}M^2_B+q^0 M_B=\frac{1}{2}(s-m^2_A-M^2_B)+\frac{1}{4}M^2_B+m^2_A,
\end{align}
where we have made use of Eq.~(\ref{q0rest}).

As can be seen in Eqs.~(\ref{tIS1}) and (\ref{tIS02}), we need the two-body $t$-matrices describing the $D^*D^*$ interaction for different isospin and spin configurations. This input is obtained following Ref.~\cite{Dai:2021vgf}, where the Bethe-Salpeter equation is solved using as kernel an amplitude obtained from effective field theories describing the interaction between two-vectors. This latter amplitude is constituted by several contributions, including that coming from a $D^*D^*\to D^*D^*$ contact term, from vector exchange in the t-channel as well as from box diagrams in which $D^*D^*\to D^* D\to D^*D^*$ by exchanging pions (in this latter case, a Gaussian form factor $e^{[(q^0)^2-\vec{q}^{\,2}]/\Lambda^2}$, with $\Lambda=1200$ MeV and $q^\mu=(q^0,\vec{q})$ being the four-momentum of the exchanged pion in the first $D^*D\pi$ vertex, is introduced in each $D^*D\pi$ vertex when integrating over $d^3q$). These amplitudes are projected on s-wave, as well as on spin, and then summed, producing an amplitude which is used to solve the Bethe-Salpeter equation. As can be seen in Ref.~\cite{Dai:2021vgf}, the $D^*D^*$ interaction with isospin 0 and $j^P=1^+$ (with $j$ being the spin of the $D^*D^*$ system) forms a bound state close to the $D^*D^*$ threshold. In particular, varying the cut-off $q_\text{max}$ from 450 MeV to 650 MeV, the mass (width) of the bound state changes from $\sim$4011 MeV to $\sim$3973 MeV  ($\sim$ 29 MeV to $\sim$100 MeV). As a consequence of the two $D^*$'s being identical particles, in case of isospin 0 but $j=0$, $2$, no states are found, while for isospin 1 there is no state with $j=1$ and the interaction for $j=0,2$ is repulsive. Thus, Eqs.~(\ref{tIS1}) and (\ref{tIS02}) simplify to
\begin{align}
t^{(1/2,0)}_1&=\frac{1}{4}t^{(0,1)}_{D^*D^*},\nonumber\\
t^{(1/2,1)}_1&=\frac{1}{16}\Big[5 t^{(1,2)}_{D^*D^*}+4t^{(1,0)}_{D^*D^*}+t^{(0,1)}_{D^*D^*}\Big],\nonumber\\
t^{(1/2,2)}_1&=\frac{1}{16}\Big[9 t^{(1,2)}_{D^*D^*}+t^{(0,1)}_{D^*D^*}\Big].
\end{align}
Note that while the input for angular momentum $J\equiv J_{AB}=s_{AB}=0$ is attractive, since it involves the $D^*D^*$ two-body $t-$matrix with isospin 0 and spin 1, there is some repulsion in the input for $J=1,\,2$ from the $D^*D^*$ two-body $t$-matrices in isospin 1 and spins 0,2. However, if the attraction in the $D^*D^*$ system overcomes such repulsion, we might find states for $J=1,2$ as well. 

A final comment, before presenting the results, is in order. In Ref.~\cite{Dai:2021vgf}, when dealing with identical particles, the so-called unitary normalization was used. Within this normalization a factor $1/\sqrt{2}$ is introduced in the $|D^*D^*\rangle$ ket to avoid double counting of contributions in the intermediate states when iterating the kernel of the Bethe-Salpeter equation. This, however, implies that when calculating the three-body $T$ matrix as $T=T_1+T_2$, a factor two must be included in $T_1$ and $T_2$. Thus, since we follow Ref.~\cite{Dai:2021vgf} to get the $D^*D^*$ two-body t-matrices, the three-body $T$-matrix must be obtained as $T=2(T_1+T_2)=4T_1$.

\section{Results}\label{res}
\begin{figure}[h!]
\centering
\includegraphics[width=0.45\textwidth]{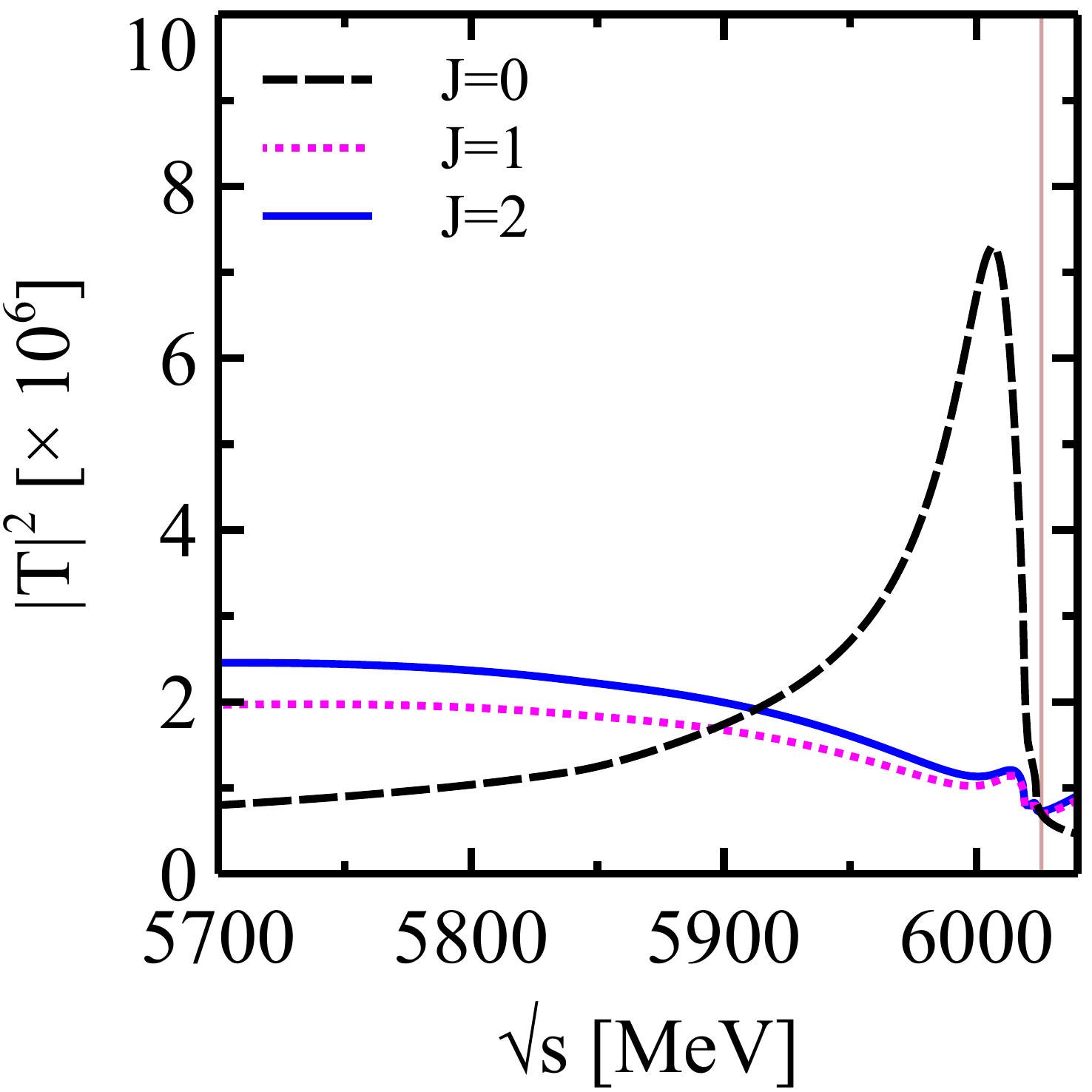}
\caption{Modulus squared of the three-body $T$-matrix as a function of $\sqrt{s}$ for $J^P=0^-$, $1^-$, $2^-$. The results correspond to a value of $q_\text{max}=450$ MeV. The vertical line indicates the three-body threshold, i.e., $3M_{D^*}$}.\label{Spin012-450}
\end{figure}

In Fig.~\ref{Spin012-450} we show the results obtained for the modulus squared of the three-body $T$-matrix of the system in isospin $1/2$ and for $J^P=0^-$, $1^-$ and $2^-$ with a cut-off  $q_\text{max}=450$ MeV. As can be seen, for $J=0$, we find a state with a mass of 6006.5 MeV, i.e, $\simeq19$ MeV below the three-body threshold, and a width of 46.6 MeV. Note that the width found is a consequence of the imaginary part present in the two-body $D^*D^*$ $t$-matrices used to solve Eq.~(\ref{T12}). This imaginary part has its origin in the $D^*D^*\to D^* D\to D^*D^*$ transition considered in Ref.~\cite{Dai:2021vgf}. We also find states for $J=1,\,2$ but since the corresponding signals are much weaker (by a factor of $\simeq 6$) than the one for $J=0$  and the former states appear smeared by the background, it  would be difficult to identify them in experimental data.  Thus, it is not very meaningful to determine their properties. Still, we provide the mass values (see Table~\ref{tablenowidth}).

\begin{figure}[h!]
 \centering
 \includegraphics[width=0.5\textwidth]{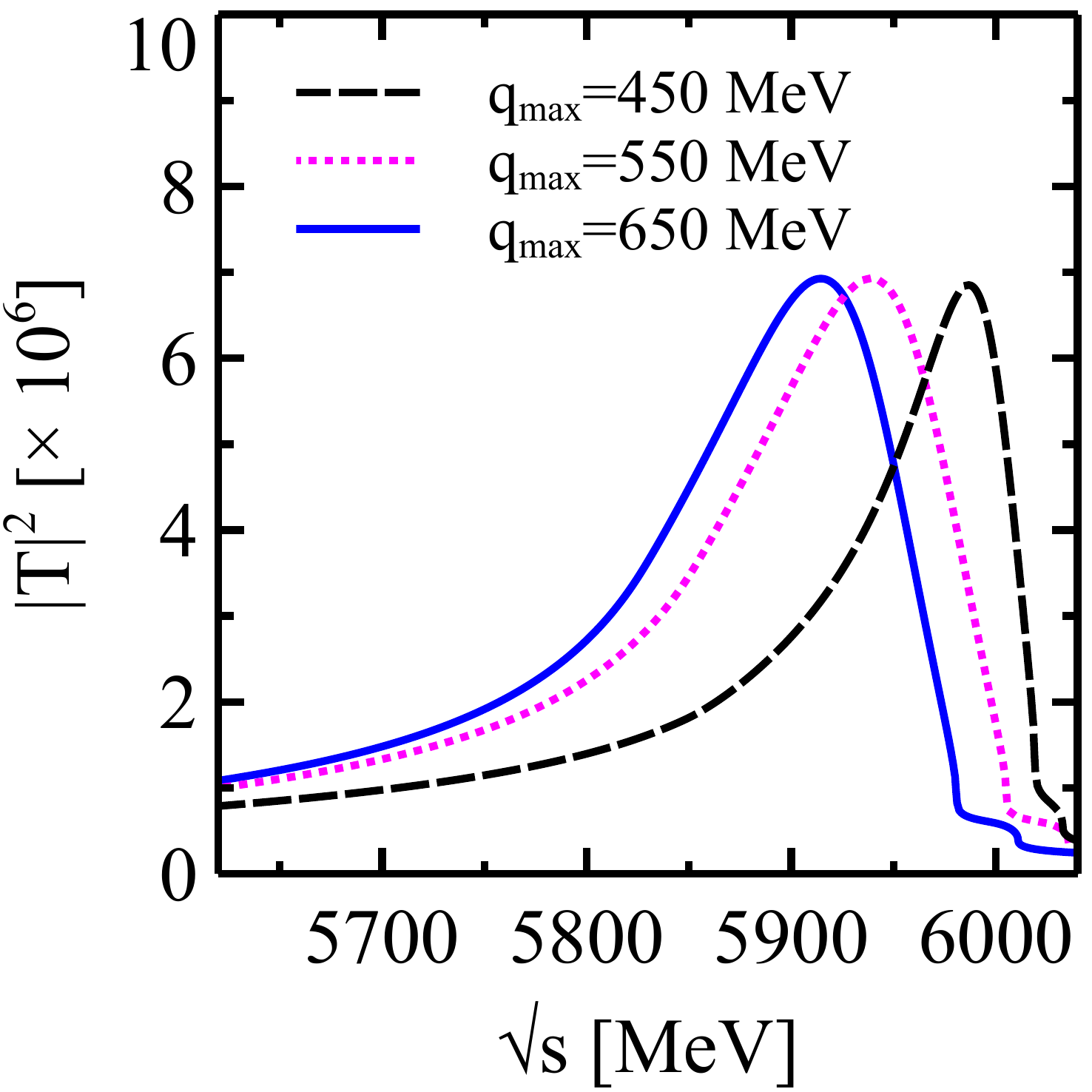}
 \caption{Modulus squared of the three-body $T$-matrix as a function of $\sqrt{s}$ for $J^P=0^-$ and $q_\text{max}=450$, $550$ and $650$ MeV.}\label{Spin012-cut}
 \end{figure}

Next, we study the uncertainty in the results produced by changing the cut-off $q_\text{max}$ used in the model of Ref.~\cite{Dai:2021vgf} when calculating the two-body $t$-matrix for the $D^*D^*$ system. In Fig.~\ref{Spin012-cut} we show the variation produced in the mass and width of the three-body state with $J=0$ for three values of $q_\text{max}=450$, $550$ and $650$ MeV. As can be seen, increasing the cut-off shifts the peak from 6006.5 MeV to 5914.5 MeV and the width increases up to 136 MeV. In Table~\ref{tablenowidth} we sumarize the masses and widths found for the states with $J=0,1,2$ when varying $q_\text{max}$. 
\begin{table}
\caption{Mass, $M$, and width, $\Gamma$, of the states found in the $D^*D^*D^*$ system with isospin $1/2$ and spin-parity $J^P=0^-$ for different values of $q_\text{max}$}\label{tablenowidth}
\begin{tabular}{c|ccc|}
&\multicolumn{3}{|c|}{$M$ ($\Gamma$) [MeV]}\\
\hline
$q_\text{max}$ [MeV]&450&550&650\\
\hline
$J=0$&~6006.5 (46.6)&~5973.5 (90.9)&~5914.5 (136.0)\\
$J=1$&~6014.1 &~5992.0 &~5954.5 \\
$J=2$&~6015.4&~5992.3 &~5954.7
\end{tabular}
\end{table}
We should however recall that consistency with the $T_{cc}$ data demanded values of the cut-off of the order of 420-450 MeV. Hence we should give credibility to the value for $q_\text{max}=450$ MeV in the Table~\ref{tablenowidth}. There is another feature worth calling the attention. The width of the state increases for more binding energy in spite of having less phase space for the decay. This is similar to what was observed in Ref.~\cite{Dai:2021vgf} for the $D^*D^*$ system and has its origin in the Weinberg compositeness condition where the coupling square of the state to the components goes as the square root of the binding energy~\cite{Weinberg:1962hj}.

\begin{figure}
\centering
\includegraphics[width=0.5\textwidth]{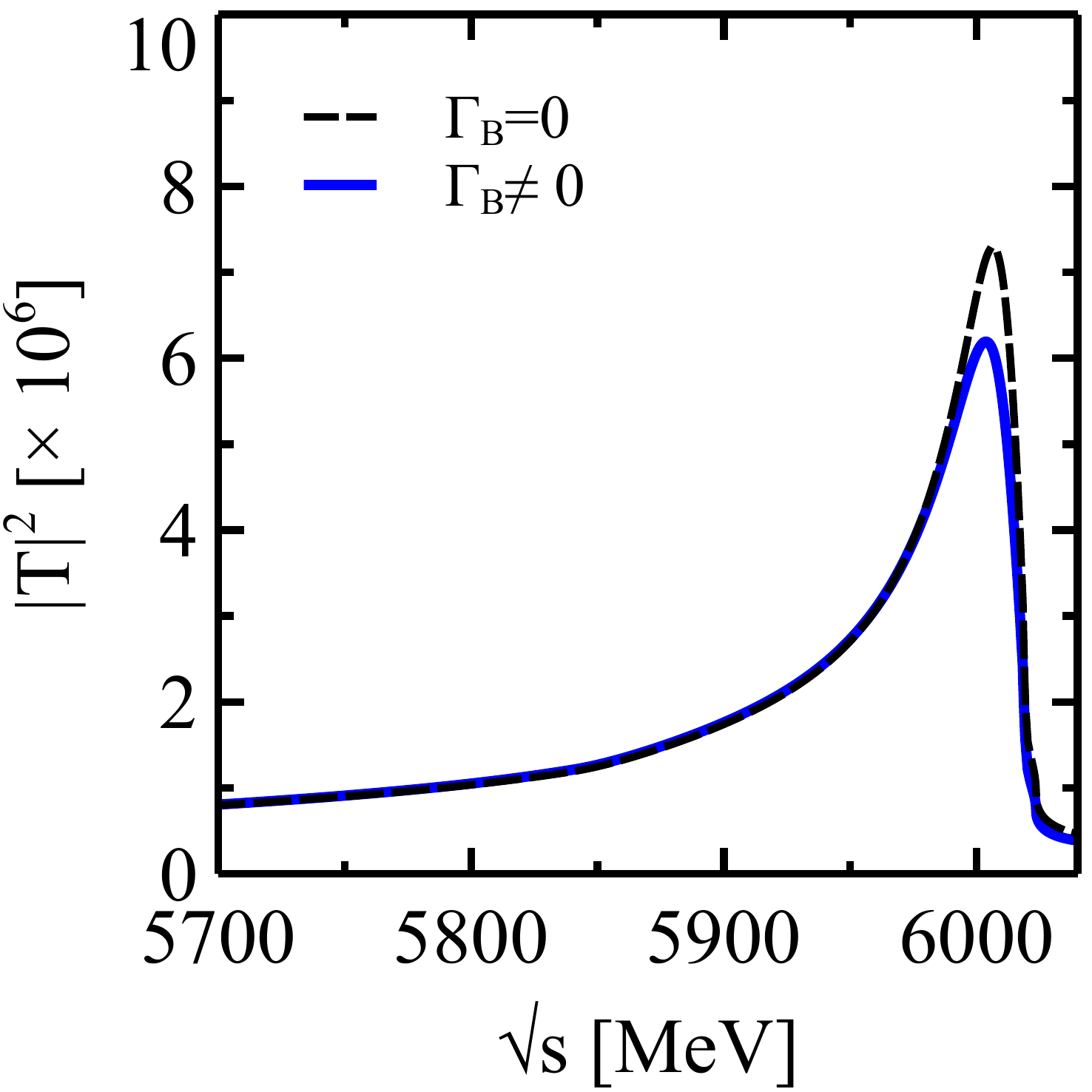}
\caption{Modulus squared of the three-body $T$-matrix as a function of $\sqrt{s}$ for $J^P=0^-$, $q_\text{max}=450$ MeV and considering the effect of the width of the cluster.}\label{Spin0width}
\end{figure}

The previous results have being obtained by neglecting the width, $\Gamma_B$, related to the cluster since $\Gamma_B<<M_B$. However, for a better estimation of the width of the three-body states found, we can evaluate the effect that the width of the cluster produces in our results. In our formalism such a width enters in the form factor written in Eq.~(\ref{FF}) and it can be incorporated by changing $M_B$ to $M_B-i\Gamma_B/2$ in Eq.~(\ref{FF}). Such a change produces a small imaginary part (when compared to the real part) for the form factor. In Fig.~\ref{Spin0width} we compare the results obtained for the modulus squared of the three-body $T$-matrix for isospin $1/2$ and $J=0$ when $\Gamma_B=0$ and considering the value of $\Gamma_B$ obtained in Ref.~\cite{Dai:2021vgf} for a cut-off $q_\text{max}=450$ MeV, which is $\Gamma_B=29.54$ MeV. As can be seen, considering the latter value of $\Gamma_B$ increases the width of the three-body state with $J=0$ by about $20\%$ for $q_\text{max}=450$ MeV and less for the other values of the cut-off. In Table~\ref{tablewidth} we summarize the results obtained when incorporating $\Gamma_B$ in the formalism. 

\begin{table}
\caption{Mass, $M$, and width, $\Gamma$, of the states found in the $D^*D^*D^*$ system with isospin $1/2$ and spin-parity $J^P=0^-$, $1^-$ and $2^-$ for different values of $q_\text{max}$ and taking into account the width of the cluster.}\label{tablewidth}
\begin{tabular}{c|ccc|}
&\multicolumn{3}{|c|}{$M$ ($\Gamma$) [MeV]}\\
\hline
$q_\text{max}$ [MeV]&450&550&650\\
$M_B$ [MeV]& 4010.7& 3997.0 & 3972.5\\
$\Gamma_B$ [MeV]&29.54 & 60.03& 99.57\\
\hline
$J=0$&~6004.5 (57.3)&~5970.9 (99.7)&~5910.8 (143.3)\\
$J=1$&~6013.6 &~5990.2&~5951.5\\
$J=2$&~6013.3&~5989.4&~5950.1
\end{tabular}
\end{table}

It is interesting to compare our results with those of Ref.~\cite{Luo:2021ggs}. In this latter work, states with a few MeV of binding energy were found for isospin $I=1/2$, spin-parity $J^P=0^-$, $1^-$, $2^-$, $3^-$. The binding is found to change with a cut off $\Lambda$ in a form factor used to regularize loops. We also find that our results depend on the cut-off $q_\text{max}$ that we use, but we should rely more on those obtained with the cut-off used to reproduce the properties of the $T_{cc}$ state, i.e., $q_\text{max}\simeq 450$ MeV. We have obtained states for $I=1/2$, $J^P=0^-$, $1^-$, $2^-$, but not $3^-$. This is a consequence of our approach since the $D^*D^*$ binds only in the $I=0$, $J^P=1^+$ configuration, hence a three-body state with $J=3$ is not possible in our model. It is interesting to see that in Ref.~\cite{Luo:2021ggs} the authors mention that there is no bound state solution with $\Lambda\simeq 1$ GeV and $J^P=3^-$, while bound states are formed for the other $J^P$ configurations. It is also mentioned that if the s-d mixing is used, then a loosely bound state for $J^P=3^-$ is obtained. Our approach is based on s-wave scattering only, hence we can say that we find the same result as in Ref.~\cite{Luo:2021ggs} when only s-waves are used.

The formalism of Ref.~\cite{Luo:2021ggs} also lead to formation of $I=3/2$ states. We cannot get such states since our cluster is isoscalar, hence we only get three body $I=1/2$ states. It is interesting what the authors of Ref.~\cite{Luo:2021ggs} mention with respect to this issue. They state that to get bound states in this case they need $\Lambda\simeq 1.8$ GeV and then conclude that since the needed cut-off $\Lambda$ is much larger than their expectation, they prefer not to view these states as good hadronic molecular states. Hence, we see that there is an agreement in the findings of both methods on the relevant cases of the bound $D^*D^*D^*$ system. 

There are also some other differences in the results of the two models: in Ref.~\cite{Luo:2021ggs} the $I(J^P)=1/2\,(0^-,\,1^-,\,2^-)$ states have similar bindings. In our case the $0^-$ state is more bound. Further, widths are not evaluated in Ref.~\cite{Luo:2021ggs}, while in our approach the widths appear automatically as a consequence of the considered $D^*D^*\to D^*D$ decays~\cite{Dai:2021vgf}. The other additional information of our approach is the strength of $|T|^2$, which is relevant to see which state has more chances to be observed in an experiment. We find that $|T|^2$ is about 6 times bigger in the case of $0^-$ than in the cases of $1^-$, $2^-$. This indicates that the $0^-$ state is the one most likely to be found in an experiment.

At this point, we find it relevant to discuss the differences in the inputs for the $D^*D^*$ interaction. We rely upon vector exchange, following the extension of the model of Ref.~\cite{Bando:1987br}, where the vector mesons are identified as the dynamical gauge bosons of hidden local symmetries. The exchange of pseudoscalars is also considered to study the $D^*D^*$ interaction in Ref.~\cite{Dai:2021vgf}, which we follow here, but only to generate the decay widths, once one realizes that its effect in the real part of the amplitudes is basically negligible as discussed in Ref.~\cite{Dias:2021upl}. Vector meson exchange is also considered in Ref.~\cite{Luo:2021ggs}, however, it is much suppressed by the form factor used, $F^2(q)=[(\Lambda^2-m^2_E)/(\Lambda^2-q^2)]^2$, where $m_E$ is the mass of the particle exchanged. For values of $\Lambda\sim0.9$ GeV, the aforementioned factor kills the vector exchange contribution in the potential by roughly a factor of 10, with the numerator of the form factor being responsible for this large reduction. We should recall that chiral Lagrangians can be obtained using vector exchange with the approach of Ref.~\cite{Bando:1987br}. In the case of $q^2=0$, and omitting $m^2_E$ in the numerator of the $F^2(q)$ mentioned above, one exactly obtains the chiral Lagrangian by exchanging the vector mesons, as shown explicitly in Ref.~\cite{Dias:2021upl}. We follow that procedure and our form factor is a sharp cut-off, $\Theta(q_\text{max}-|\vec{q}|)$, not changing the strength of the vector exchange when $q^2\to 0$.

\section{Conclusions}
We study the $D^*D^*D^*$ system considering that two of the $D^*$'s form the state found in Ref.~\cite{Dai:2021vgf}, the latter having isospin 0 and spin-parity $1^+$. By calculating the three-body scattering matrix, we find formation of bound states, with isospin $1/2$, masses $\sim 6000$ MeV and spin-parity $0^-$, $1^-$ and $2^-$. By comparing the strength of the $T$-matrices for the different spins, we find that the spin 0 state has a larger coupling to the $D^*D^*D^*$ system, thus, the signal for the spin 0 state should be more pronounced in a process in which the three states can be produced. The three states obtained have charm 3, thus, they are manifestly exotic mesons, i.e., they cannot be considered as conventional mesons formed by a quark and an antiquark. The experimental finding of these states would be a remarkable step towards the formation of a new periodic table of multimeson states with several open flavors which cannot decay into systems with a smaller number of mesons and are relatively stable.

\section*{Acknowledgements}
This work is partly supported by the Spanish Ministerio de Econom\'ia y Competitividad and European FEDER funds under Contracts No. PID2020-112777GB-I00, and by Generalitat Valenciana under contract
PROMETEO/2020/023. This project has received funding from the European Unions 10 Horizon 2020 research and innovation programme under grant agreement No. 824093 for
the ``STRONG-2020'' project. K.P.K and A.M.T thank the financial support provided by Funda\c c\~ao de Amparo \`a Pesquisa do Estado de S\~ao Paulo (FAPESP), processos n${}^\circ$ 2019/17149-3 and 2019/16924-3 and the Conselho Nacional de Desenvolvimento Cient\'ifico e Tecnol\'ogico (CNPq), grants n${}^\circ$ 305526/2019-7 and 303945/2019-2. 

\bibliographystyle{unsrt}
\bibliography{refs}

\end{document}